\documentclass[prd,floats,nopacs,twocolumn]{revtex4}
\usepackage{amsmath}
\usepackage{amssymb}
\usepackage{amsfonts}
\usepackage{dsfont}

\newcommand{\bs}[2]{{{\sf{b}}
		\phantom{]}}^{\!\!\!\!\!\!*{\kern 1.3pt}{\mbox{${\scriptstyle #1}$}}}_{\mbox{${\scriptstyle #2}$}}}
\newcommand{\as}[2]{{{\sf{a}}
		\phantom{]}}^{\!\!\!\!\!\!*{\kern 1.3pt}{\mbox{${\scriptstyle #1}$}}}_{\mbox{${\scriptstyle #2}$}}}
\newcommand{\naf}[2]{n'^{\!\!\!\!\!*{\kern 2.8pt}{\mbox{${\scriptstyle(#1)}$}}}
	_{\mbox{${\scriptstyle #2}$}}}
\newcommand{\na}[2]{n^{\!\!\!\!*{\kern 1.3pt}{\mbox{${\scriptstyle(#1)}$}}}
	_{\mbox{${\scriptstyle #2}$}}}
\renewcommand{\ap}[2]{a^{\!\!\!\!*{\kern 1.3pt}{\mbox{${\scriptstyle(#1)}$}}}
	_{\mbox{${\scriptstyle #2}$}}}
\newcommand{\ep}[1]{e^{\!\!\!\!*{\kern 1.3pt}{\mbox{${\phantom{()}}$}}}
	_{\mbox{${\scriptstyle #1}$}}}

\newcommand{\ad}[2]{a^{\!\!\!\!*{\kern 1.3pt}{\mbox{${\scriptstyle #1}$}}}_{\mbox{${\scriptstyle #2}$}}}

\newcommand{\gug}[3]{{#1}^{\!\!\!\!*{\kern 1.3pt}{\mbox{${\scriptstyle #2}$}}}_{\mbox{${\scriptstyle #3}$}}}

\newcommand{\Tr}{\mathrm{Tr}}

\newcommand {\oks}[2]{{\raise0.7ex\hbox{${\scriptstyle #1}$}\!\mathord{\left/
			{\vphantom{{1}{2}}}\right.\kern-\nulldelimiterspace}
		\!\lower0.7ex
		\hbox{${\scriptstyle #2}$}}}

\newcommand{\ii}{\mathrm{i}}

\begin{document}
	\title{
$T$ violation without complex entries in lepton mixing matrix 	
}
	
\author{A. V. Chukhnova}\email{av.chukhnova@physics.msu.ru}
\author{A. E. Lobanov}\email{lobanov@phys.msu.ru}

\affiliation {Department of Theoretical Physics, Faculty of Physics,
	Moscow State University, 119991 Moscow, Russia}

\begin{abstract}
We demonstrate that $T$ invariance can be violated even when Pontecor\-vo--Maki--Nakagawa--Sakata lepton mixing matrix is real.
We obtain a sufficient condition of $T$ violation in neutrino oscillations in matter and electromagnetic field. In the two-flavor model we derive the $T$-violating spin-flavor transition probabilities.
Then we prove that the transition probabilities for neutrino in moving medium in electromagnetic field differ from those for antineutrino in matter composed of antiparticles and electromagnetic field only in the sign of the $T$-violating term.

\end{abstract}

\maketitle

\section{Introduction}
It is well known that the physical effects discovered nowadays can be described by Lorentz-invariant theories \cite{pdg2020}. According to the $CPT$ theorem, when the charge conjugation, parity inversion and time reversal are performed at once, any local Lorentz-invariant theory with a Hermitian Hamiltonian remains invariant \cite{Pauli_CPT_rus,Jost_CPT}. Invariance under these transformations performed separately depends on the type of the interaction considered. For example, the electromagnetic interaction is known to be characterized by $P$ symmetry. In the weak interaction, as it was predicted theoretically \cite{LeeYang} and afterwards proved in experiment \cite{Vu1957}, $P$ invariance is broken and only $CP$ invariance is conserved. Later it was shown that not only $P$ symmetry, but also $CP$ symmetry is violated in experiments with kaons \cite{Cronin} and $B$-mesons \cite{BABAR2004,Belle2004}.

Studying the discrete symmetry transformations is crucial for understanding the evolution of the Universe, since $CP$ violation is required to explain the matter-antimatter asymmetry \cite{Saharov}. From the mathematical point of view, $CP$ violation in the quark sector is caused by the presence of complex elements in the Cabibbo--Kobayashi--Maskawa mixing matrix \cite{Kobayashi1973} and is determined by the Jarlskog invariant \cite{Jarlskog1985_1}. In order to study this effect it is convenient to use the parametrization of the mixing matrix for Dirac fermions, which contains three angles and one $CP$-violating phase. However, $CP$ violation due to the imaginary part of the mixing matrix still cannot provide the observed matter-antimatter asymmetry without including some other mechanisms (see, e.g., \cite{Fukugita1986,Trodden1999,Davidson2008}).

As an extra source of $CP$ violation one can consider the processes with neutrinos.
$CP$ violation will arise in neutrino oscillations, if the imaginary phase in the Pontecorvo--Maki--Nakagawa--Sakata lepton mixing matrix  \cite{Pontecorvo1957_rus,MNS} is proved to be nonzero.
Due to the $CPT$ theorem, this means the violation of $T$ symmetry in vacuum (see, e.g., \cite{Giunti_book}).

When the matter density and the values of electromagnetic fields are rather high, collective effects in neutrino oscillations become significant.
Therefore, it is not enough to consider the $CP$ and $T$ symmetries of the processes in vacuum.
The interaction with matter is usually described using the effective potential, which is associated with neutrino forward elastic scattering at the fermions of matter \cite{Wolfenstein1978} (see also \cite{Pal1989,Nieves1989_medium}).
To describe the interaction with the electromagnetic field one can use the method, called Furry picture \cite{Furry1951}. Since the neutrino is a neutral particle, the interaction of the neutrino with electromagnetic field should be considered as non-minimal \cite{Pauli1941}.
Moreover, one should take into account not only the diagonal magnetic moments, but also the transition magnetic and electric moments \cite{Fujikawa1980,Shrock1982}. It is important that the $CPT$ theorem is not fulfilled, when external fields and matter are present. In this case $T$ violation is not equivalent to $CP$ violation. That is, $T$ violation can arise independently from the $CP$-violating phase in the mixing matrix.

If the neutrino propagates in varying electromagnetic field or nonhomogeneous matter, then the formal symmetry under time reversal is obviously violated (see, e.g., \cite{EPJC2021}), since the initial and the final states of the system correspond to different external conditions. Since in the general case the result depends on the matter density and electromagnetic field profile, in the present paper we study the case of constant external conditions. Moreover, the initial and the final states are described using the quantum numbers, which correspond to the same operators of observables. This allows one to reveal the possibility of $T$ violation, which is caused by the type of the interaction, but is not caused by the behavior of external conditions and the geometry of the problem.

\section{Time reversal symmetry}
Recently we derived a $12$-component quasi-classical neutrino evolution equation, which depends on the neutrino proper time $\tau$
\begin{equation}\label{urev}
 \left( \ii \frac{d}{d \tau}- \mathcal{F}\right)\varPsi(\tau)=0
 \end{equation}
 \noindent and takes into account neutrino oscillations and spin rotation.
The most general form of matrix $\mathcal{F}$ for neutrinos interacting  with matter and electromagnetic field is  given in \cite{PR2020}.
In that paper we studied neutrino propagation in matter and electromagnetic field with constant characteristics. In this case, using the Backer--Campbell--Hausdorff formula \cite{Campbell1897} we obtained the spin-flavor transition probabilities as the following formal expansion
\begin{multline}\label{ver1}
W_{\alpha \rightarrow\beta}=\frac{1}{2u^{0}}\Tr \left\{e^{-\ii \tau {\mathcal{F}}}{\cal{P}}^{(\alpha)}_{0}e^{\ii \tau {\mathcal{F}}}{\cal{P}}^{(\beta)}_{0}(\gamma^{\mu}u_{\mu}+1)\gamma^{0}\right\} \\=
\frac{1}{2}\sum\limits^{\infty}_{n=0}\frac{(-\ii\tau)^{n}}{n!}\Tr \left\{D_{n}{\cal{P}}^{(\beta)}_{0}(\gamma^{\mu}u_{\mu}+1)\right\},
\end{multline}
\noindent where
\begin{equation}\label{ver2}
D_{0}={\cal{P}}^{(\alpha)}_{0},\;  D_{1}=[{\mathcal{F}},{\cal{P}}^{(\alpha)}_{0}],\; D_{2}=[{\mathcal{F}},[{\mathcal{F}},{\cal{P}}^{(\alpha)}_{0}]] ...
\end{equation}
\noindent Here $\mathcal{P}_0^{(\alpha)}$ and $\mathcal{P}_0^{(\beta)}$ are  projection operators on the states with definite polarization and flavor, which can be presented as products of a flavor projection operator and a polarization projection operator. There exist the solutions of Eq. \eqref{urev} in vacuum, which are characterized by constant values of parameter $u^\mu$ such that $u^2=1$ \cite{tmf2017,AnnPhys}. Therefore, quantum numbers $u^\mu$ can be interpreted as components of the neutrino $4$-velocity.

In the present paper we investigate discrete symmetries of neutrino spin-flavor transition probabilities.
$T$ violation takes place, when there are nonzero terms of odd power in $\tau$ in the power expansion of probability \eqref{ver1}.
A nonzero value of the term containing $\tau^3$ is a sufficient condition of $T$ violation, since the term linear in $\tau$ is identically equal to zero.
For neutrino in vacuum this condition is satisfied only when the mixing matrix is complex. In the general case  the interaction with matter and electromagnetic field can lead to the presence of a nontrivial coefficient in front of $\tau^3$ even when the $CP$-violating phase in the mixing matrix is equal to zero.

When only the flavor transition probabilities are considered, $T$ violation requires that the mixing matrix contains complex entries. In particular, for this reason $T$ violation cannot be present in the two-flavor model for neutrino in matter at rest (see, e.g., \cite{Naumov1992,Akhmedov2001,Petcov2018}). However, in these papers a specific situation is studied, when the neutrino helicity is an integral of motion. It means that transitions between the states with different helicities are not possible. Therefore, neutrino evolution in matter at rest is completely determined by the neutrino mixing matrix.
	The situation becomes more complicated when the neutrino helicity is not conserved. 
In particular, when neutrino propagates in electromagnetic field (see, e.g., \cite{VVO1986}) or moving matter \cite{Lobanov2001}, the transitions with the change of helicity are possible. As a result,  $T$ invariance of the transition probabilities cannot be proved in the two-flavor model. 

Using Eq. \eqref{ver1} to study all the possible situations with $T$ violation is a rather complicated task.
For this reason we will work in the two-flavor model, where the mixing matrix is real and is determined by one mixing angle. That is, we consider the case, when the $T$ invariance breaking can be caused only by the influence of external conditions.

In \cite{PR2020} we obtained an explicit form of the spin-flavor transition probabilities in the two-flavor model for two cases, which can be solved analytically: either neutrinos propagate in dense moving matter, or neutrinos propagate in electromagnetic field. In these cases the violation of $T$ invariance is absent. In the next section we will demonstrate that $T$ symmetry can be violated even in the two-flavor model, if the influence of both matter and field is considered. If the neutrino interacts with matter via neutral currents only, and the transition magnetic and electric moments are neglected, then the spin-flavor transition probabilities can be obtained in an explicit form. 
Note that taking into account the transition moments and the interaction via charged currents makes the problem mathematically more complicated, but cannot restore $T$ invariance.

\section{Neutrino in matter and electromagnetic field}
In the most simple case, when neutrino interacts with matter via neutral currents, and the transition moments are not taken into account, matrix $\mathcal{F}$, which defines the evolution equation \eqref{urev}, takes the form
\begin{multline}\label{QQ1}
\mathcal{F} =
{\mathds{M}}+
\frac{1}{2}(f^{({\mathrm N})}u){\mathds I}+\frac{1}{2}{R}_{{\mathrm N}}{\mathds I}\gamma^{5}
\gamma^{\sigma}s_{\sigma}^{({\mathrm N})}
\gamma^{\mu}u_{\mu} \\
-\mathds{M}_d
\gamma^{5}\gamma^{\mu}{\,}^{\star\!\!}F_{\mu\nu}u^{\nu}.
\end{multline}
\noindent Here $\mathds{I}$ is an identity matrix, $\mathds{M}$ is the mass matrix, and $\mathds{M}_d$ is the matrix of diagonal neutrino magnetic moments.  We assume multiplying the first two terms by  a unity matrix from the algebra of Dirac matrices. Here $\gamma^5 = - \ii \gamma^0 \gamma^1 \gamma^2 \gamma^3$, tensor ${\,}^{\star\!\!}F_{\mu\nu} = -\frac{1}{2} e^{\mu\nu\rho\lambda} F_{\rho \lambda}$ is dual to the electromagnetic field tensor $F^{\mu\nu}$, and $e^{0123}=1$. The interaction with matter is determined by effective potential $f^{(\mathrm{N})}_\mu$. In Eq. \eqref{QQ1} we use the following notations
\begin{equation}\label{t4}
\begin{array}{l} \displaystyle

s^{({\mathrm N})}_\mu\!=\!\frac{u_{\mu}(f^{(\mathrm{N})}u)-f^{(\mathrm{N})}_\mu}{R_{\mathrm{N}}}, \quad {R}_{{\mathrm N}}\!=\!{\sqrt{(f^{(\mathrm{N})} u)^2 - (f^{(\mathrm{N})})^2}}.
\end{array}
\end{equation}

Within the Standard Model  $f^{(\mathrm{N})}_\mu$ 	is defined by the $4$-vectors of current $j_{\mu}^{{(i)}}$ and polarization $\lambda_{\mu}^{(i)}$ of the matter fermions of type $(i)$ (see, e.g., \cite{izvu2017})
	\begin{equation}\label{l12-14}
	\begin{array}{l} \displaystyle
	f_\mu^{(\mathrm N)}=\sqrt{2}{G}_{{\mathrm F}}\!\sum\limits_{i}\!\!
	\Big({j_\mu^{{(i)}}}
	\big(T^{(i)}-2Q^{(i)}\sin^{2}
	\theta_{\mathrm{W}}\big)-
	{\lambda_\mu^{{(i)}}}T^{(i)}\Big),
	\end{array}
	\end{equation}
	\noindent where $G_{\mathrm F}$ is the Fermi constant, $Q^{(i)}$ is the electric charge in the units of the positron charge, $T^{(i)}$ is the weak isospin projection.
The matrix of diagonal magnetic moments $\mathds{M}_d$ for the Standard Model neutrinos in the first approximation can be expressed as a product of the mass matrix and coefficient $\mu_0$, which is given by the relation \cite{Fujikawa1980,Shrock1982}
\begin{equation}\label{mu0}
\mu_0 = \frac{3 e \mathrm{G_F}}{8\sqrt{2}\pi^2}.
\end{equation}

When matrix \eqref{QQ1} commutes with the mass matrix, that is $[\mathds{M}, \mathcal{F}]=0$, the neutrino wave function is a linear combination of the wave functions of neutrino mass eigenstates. Therefore, the effective mixing angle is equal to its vacuum value. This fact makes it possible to obtain an explicit analytical solution of the problem. We will use the two-flavor model, since, in this case, the mixing matrix is real and the resulting $T$ violation is due to spin rotation effect only.

We obtain the solution of equation \eqref{urev} using the evolution operator, which has the form
\begin{equation}
 U(\tau) = e^{-\ii \mathcal{F} \tau}.
\end{equation}
\noindent Since in the two-flavor model the mass matrix in the mass representation is diagonal, it can be expressed in the terms of the Pauli matrices as follows
\begin{equation}
\mathds{M}= \frac{1}{2}\left(m_1+m_2 + \sigma_3 (m_1-m_2)\right).
\end{equation}
The matrix of diagonal magnetic moments takes the form
	\begin{equation}\label{Mmagn}
	\mathds{M}_d= \frac{1}{2}\left(\mu_{1}+\mu_{2} + \sigma_3 (\mu_{1}-\mu_{2})\right),
\end{equation}
\noindent where $\mu_{1}$ and $\mu_{2}$ are the magnetic moments, which are approximately equal to $\mu_0 m_{1}$ and $\mu_0 m_{2}$.
\noindent Therefore, the evolution operator is given by the relation
	\begin{multline}\label{res1z}
U(\tau)\!=\!\frac{1}{4}\sum\limits_{\zeta=\pm 1}\!\!\sum\limits_{\phantom{1} k=1,2}\!\!\exp\left\{\!-\ii\tau \left[m_k+(f^{(\mathrm{N})}u)-\frac{1}{2}\zeta R_k\right]\!\right\}  \\
\times \Big(1- (-1)^k \sigma_{3}\Big)\big(1-\zeta\gamma^{5}\gamma_{\mu}s^{\mu}_{k}\big),
\end{multline}
\noindent where the indices $k=1,2$ correspond to the neutrino mass states. Here the following notations are introduced
\begin{equation}\label{ss}
\begin{array}{l}\displaystyle
	s^{\mu}_{k}= (u^{\mu}(f^{(\mathrm{N})}u)-f^{(\mathrm{N}) \mu}-2\mu_{k}{}^{\star\!\!}F^{\mu\nu}u_{\nu})/R_k, \\ [6pt]\displaystyle
 R_{k}=\Big((f^{(\mathrm{N})}u)^{2}-(f^{(\mathrm{N})})^{2} \\ \displaystyle
\phantom{R_k =}	+ 4\mu_{k}^{2}\,u^{\mu}{}^{\star\!\!}F_{\mu\alpha}\!
{}^{\star\!\!}F^{\alpha\nu}u_{\nu} -	 4\mu_{k}f^{(\mathrm{N})}_{\mu}{}^{\star\!\!}F^{\mu\nu}u_{\nu}\Big)^{1/2}\!\!\!\!. 
\end{array}
\end{equation}

Using the evolution operator we obtain the density matrix as the function of the proper time $\tau$
\begin{multline}\label{rho4}
\rho_{\alpha}(\tau)\!=\!\frac{1}{4u^{0}}U(\tau)\big(\gamma^{\mu}u_{\mu}+1\big)\Big(1-
\gamma^{5}\gamma_\mu {s}^{(\alpha)\mu}_{0}
\Big){\mathds P}_{0}^{(\alpha)}\bar{U}(\tau),
\end{multline}
\noindent where $s_0^{(\alpha)\mu}$ is the $4$-vector of initial neutrino polarization, and $\mathds{P}_{0}^{(\alpha)}$ is the projection operator on the state with flavor $\alpha$. In Eq. \eqref{rho4} we use the notation $\bar{U}(\tau)= \gamma^0 U^\dag(\tau) \gamma^0$. Our approach is valid for arbitrary initial polarization. However,
since for ultra-relativistic neutrinos the chirality almost coincides with the helicity, now we will assume that  the initial neutrino state $\alpha$ and the final neutrino state $\beta$ are the states with definite helicity.  For such states the $4$-vector of neutrino polarization can be expressed as follows
\begin{equation}\label{t61}
{s}_{0}^{(\alpha)\mu}=\zeta_{0}^{(\alpha)}{s}^{\mu}_{sp},\,\,{s}_{0}^{(\beta)\mu}=\zeta_{0}^{(\beta)}{s}^{\mu}_{sp},
\,\, {s}^{\mu}_{sp}=
\{|{\bf u}|,u^{0}{\bf u}/|{\bf u}|\},
\end{equation}
\noindent where $\zeta_0^{(\alpha)},\zeta_0^{(\beta)}=1$ correspond to right-handed neutrino, and $\zeta_0^{(\alpha)},\zeta_0^{(\beta)}=-1$ correspond to left-handed neutrino. In the two-flavor model the projection operators on the states with definite flavor can be written in the flavor representation in the form
	\begin{equation}\label{nM_mas}
\mathds{P}^{(\alpha)}_{0} =   \frac{1}{2}(1 +\xi_{0}^{(\alpha)}\sigma_3), \quad \mathds{P}^{(\beta)}_{0} =   \frac{1}{2}(1 +\xi_{0}^{(\beta)}\sigma_3).
\end{equation}
\noindent Here, according to the tradition, $\xi_0^{(\alpha)}, \xi_{0}^{(\beta)}= 1$ for neutrino with electron flavor, and $\xi_0^{(\alpha)}, \xi_{0}^{(\beta)} = -1$ otherwise. Of course, since we do not take into account the interaction via charged currents, transitions between muon flavor and tauon flavor states can also be considered in such a manner. As it is well known, to convert any operator from the flavor representation to the mass representation one needs to use the mixing matrix, which in the two-flavor model can be parameterized using one mixing angle $\theta$ (see, e.g., \cite{Giunti_book}).
In the mass representation the flavor projection operators take the form
\begin{equation}
	\begin{array}{l}\displaystyle
	\mathds{P}^{(\alpha)}_0 = \frac{1}{2}\left( 1 + \xi_0^{(\alpha)}( - \sigma_1 \sin\theta + \sigma_3 \cos\theta) \right),\\[6pt] \displaystyle
		\mathds{P}^{(\beta)}_0 = \frac{1}{2}\left( 1 + \xi_0^{(\beta)}(- \sigma_1 \sin\theta + \sigma_3 \cos\theta) \right).
		\end{array}
\end{equation}

Then the probability of the transition from the state with flavor $\alpha$ and polarization $\zeta_0^{(\alpha)}$ to the state with flavor $\beta$ and polarization $\zeta_0^{(\beta)}$ is given by the relation \cite{Neumann1927,Landau1927}
\begin{equation}\label{ver}
W_{\alpha \rightarrow\beta}=\Tr\left\{\rho_{\alpha}(\tau)\rho^{\dag}_{\beta}
(\tau=0)\right\}.
\end{equation}
Eq. \eqref{ver} contains the direct product of the Dirac matrices and flavor matrices. It is well-know that a trace of a direct product of arbitrary matrices $A$ and $B$ can be calculated as $\Tr(A\otimes B) = \Tr A \cdot \Tr B$.
Using this relation, we obtain all the spin-flavor transition probabilities. In the most general form the transition probabilities can be presented as follows
\begin{equation}\label{nl31}
\begin{array}{l}\displaystyle
W_{\alpha \rightarrow\beta} =\frac{1+\xi_0^{(\alpha)}\xi_0^{(\beta)}}{2}\frac{1+\zeta_0^{(\alpha)}
	\zeta_0^{(\beta)}}
{2}W_1  \\  \displaystyle
\phantom{W_{\alpha \rightarrow\beta}}
+\frac{1+\xi_0^{(\alpha)}\xi_0^{(\beta)}}{2}\frac{1-\zeta_0^{(\alpha)}\zeta_0^{(\beta)}}{2}W_2 \\  \displaystyle
\phantom{W_{\alpha \rightarrow\beta}} + \frac{1-\xi_0^{(\alpha)}\xi_0^{(\beta)}}{2}\frac{1+\zeta_0^{(\alpha)}\zeta_0^{(\beta)}}{2}W_3 \\ \displaystyle
\phantom{W_{\alpha \rightarrow\beta}}
+ \frac{1-\xi_0^{(\alpha)}\xi_0^{(\beta)}}{2}\frac{1-\zeta_0^{(\alpha)}\zeta_0^{(\beta)}}{2}W_4,
\end{array}
\end{equation}
\noindent where
\begin{equation}\label{1nl32}
\!\begin{array}{l}
\displaystyle W_1=\frac{1}{8}\bigg\{
\big(1-\xi_0^{(\alpha)}\cos2\theta\big)^{2}\big(1+(s_{2}s_{0})^{2}\big) \\
+ \big(1+\xi_0^{(\alpha)}\cos2\theta\big)^{2}\big(1+(s_{1}s_{0})^{2}\big) \\
+\big(1-\xi_0^{(\alpha)}\cos2\theta\big)^{2}\big(1-(s_{2}s_{0})^{2}\big)\cos\tau R_{2} \\ \displaystyle
+ \big(1+\xi_0^{(\alpha)}\cos2\theta\big)^{2}\big(1-(s_{1}s_{0})^{2}\big)\cos\tau R_{1}
\\
+\sin^{2} 2\theta \Big[\big(1-\zeta_0^{(\alpha)}\!(s_{2} s_{0})\big)\big(1-\zeta_0^{(\alpha)}\!(s_{1} s_{0})\big)
\cos\Delta^{(\!-\!)}\\
+\big(1+\zeta_0^{(\alpha)}\!(s_{2} s_{0})\big)\big(1+\zeta_0^{(\alpha)}\!(s_{1} s_{0})\big)
\cos {\Delta^{(\!+\!)}} \\
+\big(1-\zeta_0^{(\alpha)}\!(s_{2} s_{0})\big)\big(1+\zeta_0^{(\alpha)}\!(s_{1} s_{0})\big)
\cos {\Phi^{(\!-\!)}} \\
+\big(1+\zeta_0^{(\alpha)}\!(s_{2} s_{0})\big)\big(1-\zeta_0^{(\alpha)}\!(s_{1} s_{0})\big)
\cos\Phi^{(\!+\!)} \Big]\bigg\},
\end{array}
\end{equation}
\vspace*{-15pt}
\begin{equation}\label{2nl32}
\!\begin{array}{l}\displaystyle
W_2= \frac{1}{8}\bigg\{
\big(1-\xi_0^{(\alpha)}\cos2\theta\big)^{2}\big(1-(s_{2}s_{0})^{2}\big) \\
+\big(1+\xi_0^{(\alpha)}\cos2\theta\big)^{2}\big(1-(s_{1}s_{0})^{2}\big)\\
-\big(1-\xi_0^{(\alpha)}\cos2\theta\big)^{2}\big(1-(s_{2}s_{0})^{2}\big)\cos\tau R_{2}   \\
-\big(1+\xi_0^{(\alpha)}\cos2\theta\big)^{2}\big(1-(s_{1}s_{0})^{2}\big)\cos\tau R_{1}\!\\
-\sin^{2}2\theta\;\big((s_{2}s_{1})+(s_{2} s_{0})(s_{1} s_{0})\big) \\
\times \Big[\cos\Delta^{(\!-\!)} + \cos {\Delta^{(\!+\!)}} -\cos {\Phi^{(\!-\!)}} - \cos \Phi^{(\!+\!)} \Big]\\
+ \zeta_{0}^{(\alpha)}\sin^{2}2\theta \;e_{\mu\nu\rho\lambda}u^{\mu}s_{0}^{\nu} s_{1}^{\rho} s_{2}^{\lambda} \\
\times \Big[\sin \Delta^{(\!-\!)} + \sin {\Delta^{(\!+\!)}} -\sin {\Phi^{(\!-\!)}} - \sin \Phi^{(\!+\!)} \Big]\bigg\},
\end{array}
\end{equation}
\vspace*{-15pt}
\begin{equation}\label{3nl32}
\!\begin{array}{l}\displaystyle
W_3= \frac{1}{8}\sin^{2}2\theta
\bigg\{
\big(1+(s_{2}s_{0})^{2}\big)+
\big(1+(s_{1}s_{0})^{2}\big)\\
+\big(1-(s_{2}s_{0})^{2}\big)\cos\tau R_{2}
+\big(1-(s_{1}s_{0})^{2}\big)\cos\tau R_{1}\\
-\Big[\big(1-\zeta_0^{(\alpha)}\!(s_{2} s_{0})\big)\big(1-\zeta_0^{(\alpha)}\!(s_{1} s_{0})\big)
\cos\Delta^{(\!-\!)} \\
+\big(1+\zeta_0^{(\alpha)}\!(s_{2} s_{0})\big)\big(1+\zeta_0^{(\alpha)}\!(s_{1} s_{0})\big)
\cos {\Delta^{(\!+\!)}} \\
+\big(1-\zeta_0^{(\alpha)}\!(s_{2} s_{0})\big)\big(1+\zeta_0^{(\alpha)}\!(s_{1} s_{0})\big)
\cos {\Phi^{(\!-\!)}} \\
+\big(1+\zeta_0^{(\alpha)}\!(s_{2} s_{0})\big)\big(1-\zeta_0^{(\alpha)}\!(s_{1} s_{0})\big)
\cos \Phi^{(\!+\!)} \Big]\bigg\},
\end{array}
\end{equation}
\vspace*{-15pt}
\begin{equation}\label{4nl32}
\!\begin{array}{l}\displaystyle
W_4= \frac{1}{8}\sin^{2}2\theta \bigg\{
\big(1-(s_{2}s_{0})^{2}\big)+
\big(1-(s_{1}s_{0})^{2}\big)\\
-\big(1-(s_{2}s_{0})^{2}\big)\cos\tau R_{2}
-\big(1-(s_{1}s_{0})^{2}\big)\cos\tau R_{1}\\
+\big((s_{2}s_{1})+(s_{2} s_{0})(s_{1} s_{0})\big) \\ \times
\Big[\cos \Delta^{(\!-\!)}\!+\!\cos{\Delta^{(\!+\!)}}\!-\!\cos{\Phi^{(\!-\!)}}\!\!-\!\cos \Phi^{(\!+\!)} \Big] \\
-\zeta_{0}^{(\alpha)}\,e_{\mu\nu\rho\lambda}u^{\mu}s_{0}^{\nu} s_{1}^{\rho} s_{2}^{\lambda} \\ \times
\Big(\sin\Delta^{(\!-\!)}\!+\!\sin {\Delta^{(\!+\!)}}\!-\!\sin {\Phi^{(\!-\!)}}\!-\!\sin \Phi^{(\!+\!)} \Big)\bigg\}.
\end{array}
\end{equation}
\noindent Here
\begin{equation}
\begin{array}{l}
\Delta^{(\!\pm\!)} = \tau \left( (m_2-m_1) \pm (R_2- R_1)/2\right),\\
\Phi^{(\!\pm\!)} = \tau \left( (m_2-m_1) \pm (R_2 + R_1)/2\right).
\end{array}
\end{equation}
\noindent Contrary to the vacuum case, the expressions for the transition probabilities for neutrino in matter and electromagnetic field are not $T$-even. The term, which violates $T$ invariance, arises in the probabilities of transitions with the change of helicity $W_2$ and $W_4$. This term is proportional to the value, which in the laboratory reference frame can be expressed in the three dimensional form as follows
	\begin{equation}\label{Tnat}
e_{\mu\nu\rho\lambda}u^{\mu}s_{0}^{\nu} s_{1}^{\rho} s_{2}^{\lambda} = 2 \frac{\mu_{2}-\mu_{1}}{R_1 R_2}  \frac{1}{|{\bf u}|} ([{\bf u} \times {\bf f}] \cdot (u_0 {\bf B} - [{\bf u}\times{\bf E}])).
\end{equation}
\noindent Here ${\bf B}$ is the magnetic induction vector, ${\bf E}$ is the electric field strength, ${\bf f}$ is the spacial part of the $4$-vector $f^{(\mathrm{N})}_{\mu}$, which is defined by the matter velocity and polarization. Obviously, the $T$-violating term vanishes when any two of the three vectors ${\bf u}$, ${\bf f}$ and $(u_0 {\bf B} - [{\bf u}\times{\bf E}])$ are collinear. That is, the probabilities are not $T$ invariant, when the external conditions are characterized by two different preferred spacial directions. In particular, when we consider unpolarized matter at rest, the expressions for all the spin-flavor transition probabilities are $T$-invariant.

Thus, we demonstrated that  $T$ violation arises even in the two-flavor model when neutrino propagates in matter and electromagnetic field. In this case $T$ violation is caused by nontrivial external conditions, but not by complex entries in the mixing matrix.
Moreover, in the case considered above the interaction with matter is determined by the neutral currents only. It is well-known that for neutrino in matter at rest neutral currents do not contribute to the transition probabilities. However, when the electromagnetic field is present, even neutrino interaction with moving matter via neutral currents results in $T$-violating terms in the transition probabilities.
  It should be noted that $T$ violation is absent for neutrino mass states, which propagate in matter and electromagnetic field \cite{arlomur}. It means that it is not the spin rotation of neutrino, but the correlations between spin rotation and flavor oscillations that lead to T violation in our case.

\section{The symmetry properties of the transition probabilities}
Now let us discuss the discrete symmetries of the transition probabilities in detail.
	We obtained the spin-flavor transition probabilities in the quasi-classical approximation.
	Within the quasi-classical approach it is usually assumed that $T$-reversal operation corresponds to the change of the sign of time variable, which in our case is the neutrino proper time $\tau$. In quantum field theory parity inversion leads to the change of the polarization of particles in the transition probabilities (see, e.g. \cite{pdg2020}). So, we can assume that in our results it corresponds to the change of the signs of $\zeta_0^{(\alpha)}$ and $\zeta_0^{(\beta)}$, which define the neutrino initial and final polarization.
The operation of charge conjugation can be performed in the usual way \cite{Kramers} (see also \cite{LL_tom4}) for solutions of the quantum neutrino evolution equation, which was derived in \cite{PR2020}. This operation means replacing neutrino with antineutrino, but does not include any transformations of the background field and matter. After this operation we obtain another system, which is not physically equivalent  to the initial one.
 In the quantum neutrino evolution equation $C$ conjugation changes the chiral projection operator, the signs of matter potential and the signs of magnetic moments. In the three-flavor model neutrino mass matrix and flavor projection operator are transformed to their complex conjugated matrices, but here we consider the two-flavor model, where these matrices are real.
Once again we emphasize that the resulting equation describes antineutrino propagating in the same medium and electromagnetic field. 
Charge conjugation results in the following replacements in the transition probabilities. The signs of all magnetic moments are changed. Changing the chiral projection operator together with the sign of matter potential in the equation leads to a more complicated rule for the sign of matter potential in the probabilities. When the potential is included in the neutrino polarization vectors $s_k^\mu$,  the sign of the potential does not change, otherwise it should be changed.

Naturally, the probabilities obtained in this paper are not invariant under $CPT$ transformation defined for quantized neutrino wave function. This does not contradict general principles, since the model under investigation is actually a theory with Lorentz-violation \cite{Caroll1990,Kostelecky1997, Coleman1999}. Therefore, the $CPT$ theorem is not applicable.
Note that in the quantum field theory $C$, $P$ and $T$ operations are defined for creation and annihilation operators of quantized fields (see, e.g. \cite{LL_tom4}), but not for external classical background, which in our case are $F^{\mu\nu}$ and $f_\mu^{(\mathrm{N})}$.
Therefore, generalizing these transformations to a model with external fields cannot be performed in a straightforward manner in contrast to what is suggested in \cite{DPP} for generalizing Lorentz-invariance.

Since probabilities of neutrino and antineutrino transitions in matter and electromagnetic field are obviously different, let us now analyze neutrino propagation in matter composed of antiparticles. The potentials of neutrino interaction with matter are due to forward elastic scattering by the fermions of medium and depend on the matter velocity and polarization. 
After deriving the potentials of neutrino interaction with medium composed of antiparticles similarly to \cite{PR2020} we obtain the following result. Replacing the particles of the medium by the corresponding antiparticles with inverted polarization means changing the sign of the effective potential everywhere in the transition probabilities. Note that this procedure can be performed not only for the potential of interaction with matter via neutral currents, but for the potential of interaction via charged currents, too.

Now let us discuss antineutrino propagating in matter composed of antiparticles.
To study antineutrino propagation we need to change the sign of the neutrino polarization  $\zeta_0^{(\alpha)}\rightarrow -\zeta_0^{(\alpha)}$ and $\zeta_0^{(\beta)}\rightarrow -\zeta_0^{(\beta)}$.
After we change the sign of matter potential $f_\mu^{(\mathrm{N})}$ to describe propagation in matter composed of antiparticles, the expressions for transition probabilities do not differ from the initial ones.
 That is, we come to an obvious result: left-handed neutrinos propagate in the matter composed of particles in the same way, as right-handed neutrinos do in the matter composed of antiparticles. 
 
Let us now consider a neutrino interacting only with electromagnetic field. To obtain a description of antineutrino propagation one needs to change the signs of the neutrino electromagnetic characteristics, i.e the magnetic moments.
The same replacements should be performed in the transition probabilities obtained in \cite{PR2020}, where both the diagonal and the transition magnetic moments are taken into account. From the explicit form of these transition probabilities one can see that left-handed neutrinos also propagate in the electromagnetic field in the same way as right-handed antineutrinos.
It should be noted that for neutrino in matter similar result is quite expected, since even for one flavor the weak interaction is $CP$-even. However, although the electromagnetic interaction is $P$-even for one flavor, to obtain the same probabilities we still need to change neutrino helicity.

In the present paper we obtained the spin-flavor transition probabilities for neutrino, which propagates in matter and electromagnetic field. 
Here we take into account only the diagonal magnetic moments and the interaction with matter via neutral currents. Therefore, changing neutrino to antineutrino and changing fermions of medium to their antiparticles results in the change of the signs of $s_k^\mu$ and the the signs of $\zeta_0^{(\alpha)}$ and $\zeta_0^{(\beta)}$.
In this case the transition probabilities for left-handed neutrino in ordinary matter and for right-handed antineutrino in matter composed of antiparticles are not similar anymore. To restore the initial expressions, one also needs to change the sign of the proper time $\tau$.
Thus, we obtain an interesting result. In the presence of electromagnetic field the probabilities obtained for transitions of left-handed neutrino in matter differ from those for right-handed antineutrino in matter composed of antiparticles only in the sign of the $T$-violating term. Note that this is not a direct consequence of the $CPT$-theorem, since our model is Loretz-violating and the conditions of the $CPT$-theorem are not fulfilled.

\section{Conclusion}
The model we consider is a theory with Lorentz violation and so the neutrino transition probabilities in matter and magnetic field are not invariant under $CPT$ transformation performed for neutrino. We discuss neutrino and antineutrino propagation in ordinary matter and matter composed of antiparticles. We show that in an external electromagnetic field a system of weakly interacting particles does not evolve in the same way as the system of their antiparticles.

In the two-flavor model we derive the transition probabilities for neutrino in matter and electromagnetic field and obtain $T$ violating terms.
$T$ violation implies a difference between the probabilities of transitions from the state $\alpha$ to the state $\beta$ and from the state $\beta$ to the state $\alpha$ ($W_{\alpha\rightarrow\beta}-W_{\beta\rightarrow\alpha}\neq 0$).
$T$ violation in two-flavor neutrino oscillations becomes possible due to the combined effect of matter and electromagnetic field. In this case $T$ violation is actually caused by the fact that the neutrino helicity is not conserved. Obviously, it cannot arise for two-flavor neutrino in matter at rest.

That is, in the present paper we reveal an extra source of $T$ violation, which emerges due to collective effects and is not caused by complex entries in the mixing matrix. This source of $T$ violation does not require introducing new symmetries, which imply either new particles to be included into the Standard Model or new properties of the presently known particles.

\acknowledgments
{
The authors are grateful to A.\,V. Borisov, I.\,P. Volobuev and V.\,Ch. Zhukovsky for fruitful discussions. A.V.C. is grateful to the Theoretical Physics and Mathematics Advancement Foundation
``BASIS'' (grant No. 19-2-6-100-1).}

\end{document}